\documentstyle[12pt]{article}
\newcommand{\be}{\begin{equation}}
\newcommand{\ee}{\end{equation}}
\newcommand{\bea}{\begin{eqnarray}}
\newcommand{\eea}{\end{eqnarray}}

\begin{document}

\title{TOPOLOGICAL UNCONSTRAINED QCD }

\author{
Victor Pervushin, \\
{\normalsize\it Joint Institute for Nuclear Research},\\
 {\normalsize\it 141980, Dubna, Russia.}
}


\maketitle

\medskip
\medskip

\begin {abstract}

"Equivalent unconstrained systems" for QCD obtained
by resolving the Gauss law are discussed.
We show that the effects of hadronization, confinement,
spontaneous chiral symmetry breaking and $\eta_0$-meson mass
can be hidden
in solutions of the non-Abelian Gauss constraint in the class of
functions of topological gauge transformations, in the form of
a monopole, a zero mode of the Gauss law, and a rising potential.

\end{abstract}


(Key-words: QCD, Gauss law, topology, monopole, zero mode,
hadronization, confinement, U(1)-problem)

\section{Introduction}

The consistent dynamic description of gauge constrained systems
was one of the most fundamental problems of theoretical physics
in the 20th century. There is an opinion that
the highest level of solving the problem of quantum description of
gauge relativistic constrained systems is the Faddeev-Popov (FP) integral
for unitary perturbation theory~\cite{fp1}. In any case, just this FP
integral was the basis to prove renormalizability of the unified theory
of electroweak interactions in papers by 't Hooft and Veltman marked by
the 1999 Nobel prize.

Another opinion is that the FP integral has only the intuitive status.
The most fundamental level of the description of gauge constrained
systems is the derivation of "equivalent unconstrained systems"
compatible with the simplest variation methods of the Newton mechanics
and with the simplest quantization by the Feynman path integral.
It was the topic of Faddeev's paper~\cite{f} " Feynman integral
for singular Lagrangians" where
the non-Abelian "equivalent unconstrained system" was obtained (by
explicit resolving the Gauss law), in order to justify the intuitive
FP path integral~\cite{fp1} by its equivalence to the Feynman path
integral. Faddeev managed to prove the equivalence of the Feynman integral
to the FP one only for the scattering amplitudes~\cite{f} where
all particle-like excitations of the fields are on their mass-shell.
However, this equivalence is not proved and becomes doubtful for the cases of
bound states, zero modes and other collective phenomena
where these fields are off their mass-shell. It is just the case of QCD.
In this case, the FP integral in an arbitrary relativistic gauge can lose
most interesting physical phenomena
hidden in the explicit solutions of constraints~\cite{n,rev}.

The present paper is devoted to the derivation an
"equivalent unconstrained systems" for QCD
in the class of functions of topologically nontrivial
transformations, in order to present here a set of arguments in favor of
that physical reasons of hadronization and confinement in QCD
can be hidden in the explicit solutions of the non-Abelian constraints.

\section{Equivalent Unconstrained Systems in QED}

The Gauss law constraint is the equation for
the time component of a gauge field
\be  \label{glqed}
\frac{\delta W}{\delta A_0}=0~\Rightarrow~~~\partial_j^2A_0=
\partial_k \dot A_k +J_0
\ee
in the frame of reference with an axis of time
$l^{(0)}_{\mu}=(1,0,0,0)$.
Heisenberg and Pauli~\cite{hp} noted that the gauge
($\partial_k A^*_k \equiv 0$) is distinguished,
and Dirac~\cite{cj} constructed the corresponding ("dressed") variables
$A^*$ in the explicit form
\be \label{df}
ieA^*_k=U(A)(ieA_k+\partial_k )U(A)^{-1}~,~~~~~~~~
U(A)=\exp[ie\frac{1}{\partial_j^2 }\partial_kA_k ]~,
\ee
using for the phase the time integral of the spatial part of the
Gauss law $\partial_k\dot A_k $. The action for an equivalent
unconstrained system (EUS) for QED is derived by
the substitution of the solution of the Gauss law in terms of the "dressed"
variables into the initial action
\be\label{qed}
W_{Gauss-shell}= W^*_{l^{(0)}}(A^*,E^*)~.
\ee
The peculiarity of the "equivalent unconstrained system" for QED is
the electrostatic phenomena described by the monopole class of functions
($f(\vec x)= O(1/r), |\vec x|=r \rightarrow \infty$).

The "equivalent unconstrained system"  can be quantized by
the Feynman path integral in the form
\be \label{qedf}
Z_F[l^{(0)},J^*] =\int\limits_{ }^{ }d^2A^* d^2 E^*
\exp\left\{iW^*_{l^{(0)}} [A^*,E^*]+i\int\limits_{ }^{ }d^4x
[J^*_k \cdot A^*_k-J_0^*\cdot A_0^*]\right\}
\ee
where $J^*$ are physical sources.
This path integral depends on the axis of time $l^{(0)}_{\mu}=(1,0,0,0)$.

One supposes that the dependence on the frame ($l^{(0)}$) can be removed
by the transition from the Feynman integral of "EUS" ~(\ref{qedf})
to perturbation theory in any relativistic-invariant
gauge $f(A)=0$ with the FP determinant
\be \label{qedr}
Z_{FP}[J] =\int\limits_{ }^{ }d^4A \delta[f(A)]\Delta_{FP}
\exp\left\{iW[A]-i\int\limits_{ }^{ }d^4xJ_{\mu} \cdot A^{\mu}\right\}~.
\ee
This transition is well-known as a "change of gauge", and it
is fulfilled in two steps \\
I) the change of the variables $A^*$~(\ref{df}), and \\
II) the change of the physical sources $J^*$ of the type of
\be \label{ps}
A^*_k(A)J_k^*=U(A)\left(A_k-\frac{i}{e}\partial_k \right)U^{-1}(A) J^*_k
~\Rightarrow~A_{k}J^{k}~.
\ee
At the first step, all electrostatic monopole physical phenomena
are concentrated in the Dirac gauge factor $U(A)$~(\ref{df})
that accompanies the physical sources $J^*$.

At the second step, changing the sources~(\ref{ps}) we lose the Dirac factor
together with the whole class
of electrostatic phenomena including the Coulomb-like instantaneous bound
state formed by the electrostatic interaction.

Really, the FP perturbation theory in the relativistic gauge contains only
photon propagators with the light-cone singularities forming
the Wick-Cutkosky bound states with the spectrum  differing
\footnote{The author thanks W. Kummer who pointed out that in
Ref. \cite{kum} the difference between the Coulomb atom and
the Wick-Cutkosky bound states in QED has been demonstrated.}
from the observed one which corresponds to the instantaneous Coulomb
interaction.
Thus, the restoration of the explicit relativistic form of EUS($l^{(0)}$)
by the transition to a relativistic gauge
loses all electrostatic "monopole physics" with the Coulomb bound states.

In fact, a moving relativistic atom in QED is described by the usual boost
procedure for the wave function, which corresponds to a change of the
time axis $l^{(0)}\Rightarrow l$, i.e.,
motion of the Coulomb potential~\cite{yaf} itself
\begin{eqnarray}
{ W}_{C}
= \int d^4 x d^4 y \frac{1}{2}
J_{l}(x)
V_C(z^{\perp})
J_{l}(y) \delta(l \cdot z) \,\,\, ,
\end{eqnarray}
where
$
J_{l} = l_{\mu} J^{\mu} \,\, , \,\,
z_{\mu}^{\perp} =
z_{\mu} - l_{\mu}(z \cdot l) \,\, , \,\,
z_\mu = (x - y)_\mu  \,\, . \,\,
$
This transformation law and the relativistic covariance of EUS
in QED has been predicted by von Neumann~\cite{hp} and
proven by Zumino \cite{z} on the level of the algebra of
generators of the Poincare group.
Thus, on the level of EUS, the choice of a gauge is reduced to the choice of
a time axis (i.e., the reference frame). A time axis is chosen
to be parallel to the total momentum of a bound state, so that the
coordinate of the potential always coincides with the space of the relative
coordinates of the bound state wave function to satisfy the
Yukawa-Markov principle~\cite{ym} and the Eddington concept of
simultaneity ("yesterday's electron and today's proton do not make
an atom")~\cite{ed}.

In other words, each instantaneous bound state in QED has a proper EUS,
and the relativistic generalization of the potential model is not only
the change of the form of the potential, but sooner the change of a
direction of its motion in four-dimensional space to lie along the
total momentum of the
bound state. The relativistic covariant unitary perturbation theory
in terms of instantaneous bound states has been constructed in~\cite{yaf}.

\section{Unconstrained QCD}

\subsection{Topological degeneration and class of functions}

We consider the non-Abelian $SU_c(3)$ theory with the action functional
\be \label{u}
W=\int d^4x
\left\{\frac{1}{2}({G^a_{0i}}^2- {B_i^a}^2)
+ \bar\psi[i\gamma^\mu(\partial _\mu+{\hat
A_\mu})
-m]\psi\right\}~,
\ee
where $\psi$ and $\bar \psi$ are the fermionic quark fields.
We use the conventional notation for the non-Abelian electric
and magnetic fields
\be \label{v}
G_{0i}^a = \partial_0 A^a_i - D_i^{ab}(A)A_0^b~,~~~~~~
B_i^a=\epsilon_{ijk}\left(\partial_jA_k^a+
\frac g 2f^{abc}A^b_jA_k^c\right)~,
\ee
as well as the covariant derivative
$D^{ab}_i(A):=\delta^{ab}\partial_i + gf^{acb} A_i^c$.

The action (\ref{u}) is invariant with respect to gauge transformations
$u(t,\vec x)$
\be \label{gauge1}
{\hat A}_{i}^u := u(t,\vec x)\left({\hat A}_{i} + \partial_i
\right)u^{-1}(t,\vec x),~~~~~~
\psi^u := u(t,\vec x)\psi~,
\ee
where ${\hat A_\mu}=g\frac{\lambda^a }{2i} A_\mu^a~$.

It is well-known~\cite{fs} that the initial data of all fields are
degenerated with respect to the stationary gauge transformations
$u(t,\vec{x})=v(\vec{x})$.
The group of these transformations represents
the group of three-dimensional paths lying  on the three-dimensional
space of the $SU_c(3)$-manifold with the homotopy group
$\pi_{(3)}(SU_c(3))=Z$.
The whole group of stationary gauge transformations is split into
topological classes marked by the integer number $n$ (the degree of the map)
which counts how many times a three-dimensional path turns around the
$SU(3)$-manifold when the coordinate $x_i$ runs over the space where it is
defined.
The stationary transformations $v^n(\vec{x})$ with $n=0$ are called the small
ones; and those with $n \neq 0$
\be \label{gnl}
{\hat A}_i^{(n)}:=v^{(n)}(\vec{x}){\hat A}_i(\vec{x})
{v^{(n)}(\vec{x})}^{-1}
+L^n_i~,~~~~L^n_i=v^{(n)}(\vec{x})\partial_i{v^{(n)}(\vec{x})}^{-1}~,
\ee
the large ones.

The degree of a map
\be \label{gn2}
{\cal N}[n]
=-\frac {1}{24\pi^2}\int d^3x ~\epsilon^{ijk}~ Tr[L^n_iL^n_jL^n_k]=n~.
\ee
as the condition for
normalization  means that the large transformations
are given in the  class of functions with the spatial asymptotics
{\cal O}$(1/r)$.
Such a function $L^n_i$~(\ref{gnl}) is given by
\be \label{class0}
v^{(n)}(\vec{x})=\exp(n \hat \Phi_0(\vec{x})),~~~~~
\hat \Phi_0=- i \pi\frac{\lambda_A^a x^a}{r} f_0(r)~,
\ee
where the antisymmetric SU(3) matrices are denoted by
$$\lambda_A^1:=\lambda^2,~\lambda_A^2:=\lambda^5,~\lambda_A^3:=\lambda^7~,$$
and $r=|\vec x|$.
The function $f_0(r)$ satisfies the boundary conditions
\be \label{bcf0}
f_0(0)=0,~~~~~~~~~~~~~~
f_0(\infty)=1~,
\ee
so that the functions $L_i^n$ disappear at spatial infinity
$\sim$ {\cal O}$(1/r)$.
The functions $L_i^n$ belong to monopole-type class of
functions. It is evident that the transformed physical fields $A_i$
in~(\ref{gnl}) should be given in the same class of functions.

The statement of the problem is {\bf to construct an
equivalent unconstrained system (EUS) for the non-Abelian fields
in this monopole-type class of functions}.

\subsection{The Gauss Law Constraint}

So, to construct EUS, one should solve the non-Abelian Gauss law constraint
\cite{n,vp1}
\be \label{gaussd}
\frac{\delta W}{\delta A_0}=0~~~~~ \Rightarrow
(D^2(A))^{ac} { A_0}^c = D_i^{ac}(A)\partial_0 A_i^c+ j_0^a~,
\ee
where $j_\mu^a=g\bar \psi \frac{\lambda^a}{2} \gamma_\mu\psi$
is the quark current.

As dynamical gluon fields $A_i$ belong to a class of
monopole-type functions, we restrict ouselves to
ordinary perturbation theory around a static monopole $\Phi_i(\vec x) $
\be \label{bar}
A^c_i(t,\vec{x}) = {\Phi}^c_i(\vec{x}) + \bar A^c_i(t,\vec{x})~,
\ee
where $\bar A_i$ is a weak perturbative part
with the asymptotics at the spatial infinity
\be \label{ass1}
\hat {\Phi}_i(\vec{x})= O(\frac{1}{r}),~~~~~~~~
\bar A_i(t,\vec{x})|_{\rm asymptotics} = O(\frac{1}{r^{1+l}})~~~~(l > 1)~.
\ee
We use, as an example, the Wu-Yang monopole~\cite{wy,fn}
\be \label{wy}
\Phi_i^{WY}=
- i \frac{\lambda_A^a}{2}\epsilon_{iak}\frac{x^k}{r^2} f_1^{WY},~~~~~~~
f^{WY}_{1}=1
\ee
which is a solution of classical equations everywhere besides
the origin of coordinates.
To remove a sigularity at the origin of coordinates and regularize
its energy, the Wu-Yang monopole is changed by the
Bogomol'nyi-Prasad-Sommerfield (BPS) monopole~\cite{bps}
\be
f^{WY}_{1}~\Rightarrow~
f^{BPS}_{1}=
\left[1 - \frac{r}{\epsilon \sinh(r/\epsilon)}\right]~,~~~~~
\int\limits_{ }^{ }d^3x [B^a_i(\Phi_k)]^2 =\frac{4\pi}{g^2 \epsilon}~,
\ee
to take the limit of zero size $\epsilon~\rightarrow~ 0$ at the
end of the calculation of spectra and matrix elements.
This case gives us the possibility to obtain the phase of the
topological transformations~(\ref{class0}) in the form of
the zero mode of the covariant Laplace operator in the monopole field
\be \label{lap}
(D^2)^{ab}({\Phi_k^{BPS}})({\Phi}_0^{BPS})^b(\vec{x})=0~.
\ee
The nontrivial solution of this equation is well-known~\cite{bps};
it is given by equation~(\ref{class0}) where
\be \label{lapb}
f_0^{BPS}=\left[ \frac{1}{\tanh(r/\epsilon)}-\frac{\epsilon}{r}\right]
\ee
with the boundary conditions~(\ref{bcf0}).
This zero mode signals about a topological excitation of
the gluon system as a whole in the form of the solution ${\cal Z}^a$  of
the homogeneous equation
\be\label{zm}
(D^2(A))^{ab}{\cal Z}^b=0~,
\ee
i.e., a zero mode of the Gauss law constraint~(\ref{gaussd})~\cite{vp1,p2}
with the asymptotics at the space infinity
\be \label{ass}
\hat {\cal Z}(t,\vec{x})|_{\rm asymptotics}=\dot N(t)\hat \Phi_0(\vec{x})~,
\ee
where $\dot N(t)$ is the global variable of this topological excitation
of the gluon system as a whole.
From the mathematical point of view, this means that
the general solution of the inhomogeneous equation~(\ref{gaussd})
for the time-like component $A_0$
is a sum of the homogeneous equation~(\ref{zm})
and a particular solution
${\tilde A}_0^a$ of the inhomogeneous one~(\ref{gaussd}):
$A_0^a = {\cal Z}^a + {\tilde A}^a_0$~.

The zero mode of the Gauss constraint and the
topological variable $N(t)$ allow us to remove the topological
degeneration of all fields by the non-Abelian generalization of
the Dirac dressed variables~(\ref{df})
\be \label{gt1}
0=U_{\cal Z}(\hat {\cal Z}+\partial_0)U_{\cal Z}^{-1}~,~~~~
{\hat A}^*_i=U_{\cal Z}({\hat A}^I+\partial_i)U_{\cal Z}^{-1},~~~
\psi^*=U_{\cal Z}\psi^I~,
\ee
where the spatial asymptotics of $U_{\cal Z}$ is
\be \label{UZ}
U_{\cal Z}=T\exp[\int\limits^{t} dt'
\hat {\cal Z}(t',\vec{x})]|_{\rm asymptotics}
=\exp[N(t)\hat \Phi_0(\vec{x})]=U_{as}^{(N)}~,
\ee
and $A^I=\Phi+\bar A,\psi^I$ are the degeneration free variables
with the Coulomb-type gauge in the monopole field
\be \label{qcdg}
D_k^{ac}(\Phi)\bar A^c_k=0~.
\ee
In this case, the topological degeneration of all color fields
converts into the degeneration of only one global topological
variable $N(t)$ with respect to a shift of this variable on integers:
$(N~\Rightarrow~ N+n,~ n=\pm 1,\pm 2,...)$.
One can check~\cite{bpr} that the Pontryagin index for
the Dirac variables~(\ref{gt1}) with the
assymptotics~(\ref{ass1}),~(\ref{ass}),~(\ref{UZ}) is determined
by only the diference of the final and initial values of
the topological variable
\be \label{pont}
\nu[A^*]=\frac{g^2}{16\pi^2}\int\limits_{t_{in} }^{t_{out} }dt
\int\limits_{ }^{ }d^3x G^a_{\mu\nu} {}^*G^{a\mu\nu}=N(t_{out}) -N(t_{in})
\ee
The considered case corresponds
to the factorization of the phase factors of the topological
degeneration, so that
the physical consequences of the degeneration with respect to the
topological nontrivial initial data are determined by the gauge of
the sources of the Dirac dressed fields $A^*,\psi^*$
\be \label{tcsa}
W^*_{l^{(0)}}(A^*) + \int\limits_{ }^{ }d^4x J^{c*}A^{c*}=
W^*_{l^{(0)}}(A^I) +\int\limits_{ }^{ } d^4x J^{c*}A^{c*}(A^I)~.
\ee
The nonperturbative
phase factors of the topological degeneration can lead to
a complete destructive interference of color amplitudes~\cite{n,p2,pn}
due to averaging over all parameters of the degenerations, in particular
\be \label{conf}
<1|\psi^*|0>=<1|\psi^I|0> \lim\limits_{L \to \infty}
\frac{1}{2L} \sum\limits_{n=-L }^{n=+L } U_{as}^{(n)}(x)=0~.
\ee
This mechanism of confinement due to the interference
of phase factors (revealed by the explicit
resolving the Gauss law constraint~\cite{n}) disappears
after the change of "physical" sources $A^*J^*~\Rightarrow~A J$ (that
is called the transition to another gauge). Another gauge of the sources
loses the phenomenon of confinement, like
a relativistic gauge of sources
in QED~(\ref{ps}) loses the phenomenon of electrostatics in QED.

\subsection{Physical Consequences}

The dynamics of physical variables including the topological one
is determined by the constraint-shell action of an equivalent unconstrained
system (EUS) as a sum
of the zero mode part, and the monopole and perturbative ones
\be \label{csa}
W^*_{l^{(0)}}=W_{Gauss-shell}=W_{\cal Z}[N]+W_{mon}[\Phi_i]+W_{loc}[\bar A]~.
\ee
The action for an equivalent unconstrained system~(\ref{csa}) in the
gauge~(\ref{qcdg}) with a monopole and a zero mode has been
obtained  in the paper~\cite{bpr} following the paper~\cite{f}.
This action contains the dynamics of the topological variable in the
form of a free rotator
\be \label{ktg}
W_{\cal Z}=\int\limits_{ }^{ }dt\frac{{\dot N}^2 I}{2};~~~
I=\int\limits_{V}d^3x(D^{ac}_i(\Phi_k)\Phi^c_0)^2=
\frac{4\pi}{g^2}(2\pi)^2\epsilon~,
\ee
where $\epsilon$ is a size of the BPS monopole considered
as a parameter of the infrared regularization which disappears
in the infinite volume limit. The dependence of $\epsilon$ on
volume can be chosen so that the density of energy was finite.
In this case, the U(1) anomaly can lead to additional mass of the
isoscalar meson due to its mixing with the topological variable~\cite{bpr}.
The vacuum wave function of the topological free motion
in terms of the Ponryagin index~(\ref{pont}) takes
the form of a plane wave $\exp(i P_N \nu[A^*])$.
The well-known instanton wave
function~\cite{hooft} appears for nonphysical values of the topological
momentum $P_N=\pm i 8 \pi^2/g^2$ that points out the possible status
of instantons as nonphysical solutions with the zero energy
in Euclidean space-time
\footnote{ The author is grateful to V.N. Gribov for the discussion of
the problem of instantons in May of 1996, in Budapest.}.
In any case, such the Euclidean solutions
cannot describe the phenomena of the type of the complete destructive
interference~(\ref{conf}).

The Feynman path integral for the obtained unconstrained system in
the class of functions of the topological transformations
takes the form (see~\cite{bpr})
\be \label{qcdf}
Z_F[l^{(0)},J^{a*}] =\int\limits_{ }^{ } DN(t)
\int\limits_{ }^{ }\prod\limits_{c=1 }^{c=8 }
[d^2A^{c*} d^2 E^{c*}]
\exp\left\{iW^*_{l^{(0)}} [A^*,E^*]+i\int\limits_{ }^{ }d^4x
[J^{c*}_{\mu} \cdot A^{c*}_{\mu}]\right\}~,
\ee
where $J^{c*}$ are physical sources.

The perturbation theory in the sector of local excitations
is based  on the Green function $1/D^2(\Phi)$ as the inverse
differential operator of the Gauss law which is
the non-Abelian generalization of the Coulomb potential.
As it has been shown in~\cite{bpr}, the non-Abelian Green function
in the field of the Wu-Yang monopole
is the sum of a Coulomb-type potential and a rising one.
This means that the instantaneous quark-quark interaction
leads to spontaneous chiral symmetry breaking~\cite{yaf,fb},
goldstone mesonic bound states~\cite{yaf}, glueballs~\cite{fb,ac}, and
the Gribov modification of the asymptotic freedom formula~\cite{ac}.
If we choose a time-axis $l^{(0)}$
along the total momentum of bound states~\cite{yaf}
(this choice is compatible with the experience of QED in the description
of instantaneous bound states), we get the bilocal generalization of
the chiral Lagrangian-type mesonic interactions~\cite{yaf}.

The change of variables $A^*$ of the type of~(\ref{df})
with the non-Abelian Dirac factor
\be \label{dirqcd}
U(A)=U_{\cal Z}\exp\left\{\frac{1}{D^2(\Phi)} D_j(\Phi)\hat A_j\right\}
\ee
and the change of the Dirac dressed sources $J^*$ can remove all
monopole physics, including confinement and hadronization,
like similar changes~(\ref{df}),~(\ref{ps}) in QED (to get a
relativistic form of the Feynman path integral)
remove all electrostatic phenomena in the relativistic gauges.

The transition to another gauge faces the problem of zero
of the FP determinant $det D^2(\Phi)$ (i.e. the Gribov ambiguity~\cite{g} of
the gauge~(\ref{qcdg})). It is the zero mode of the second class
constraint. The considered example~(\ref{qcdf}) shows that
the Gribov ambiguity (being simultaneously the zero mode of the first
class constraint) cannot be removed by the change of gauge
as the zero mode is the inexorable consequence of internal dynamics, like the
Coulomb field in QED. Both the zero mode, in QCD, and the Coulomb field,
in QED, have nontrivial physical consequences discussed above,
which can be lost by the standard gauge-fixing scheme.

\section{Instead of Conclusion}

The variational methods of describing dynamic systems were created
for the Newton mechanics. All their peculiarities (including
time initial data, spatial boundary conditions $O(1/r)$, time evolution,
spatial localization, the classification
of constraints, and equations of motion in the Hamiltonian approach) reflect
the choice of a definite frame of reference distinguished by the axis of time
$l^{(0)}_{\mu}=(1,0,0,0)$. This frame determines also the
"equivalent unconstrained system" for the relativistic gauge theory.
This "equivalent system" is compatible with the simplest variational
methods of the Newton mechanics.
The manifold of frames corresponds to
the manifold of "equivalent unconstrained systems".
{\bf The relativistic invariance means that a complete set of physical states
for any "equivalent system" coincides with the one for another
"equivalent system"}~\cite{bww}.

This Schwinger's treatment of the relativistic invariance is confused
with the naive  understanding of the relativistic invariance as
{\bf independence on the time-axis of each physical state}.
The latter is not obliged, and it can be possible only for the QFT
description of local elementary excitations on their mass-shell.

For a bound state, even in QED, the dependence on the time-axis exists.
In this case, the time-axis is chosen to lie along the total momentum of
the bound state in order to get the relativistic covariant dispersion law and
invariant mass spectrum.
This means that for the description of the processes
with some bound states with different total momenta we are forced
to use also some corresponding "equivalent unconstrained systems".
Thus, a gauge constrained system can be completely covered by a set of
"equivalent unconstrained systems".
This is not the defect of the theory, but the method developed
for the Newton mechanics.

What should we choose to prove confinement and compute the
hadronic spectrum in QCD: "equivalent unconstrained systems"
obtained by the honest and direct resolving constraints, or
relativistic gauges with the lattice calculations in the Euclidean
space with the honest summing of all diagrams that lose from the
very beginning all constraint effects?

\section*{Acknowledgments}

\medskip

I thank  Profs. D. Blaschke, A. V. Efremov, E. A. Kuraev, and G. R\"oepke
for critical discussions.

\end{document}